\documentclass{PoS}
\def\frs{\phi_{e}^{S}:\phi_{\mu}^{S}:\phi_{\tau}^{S}}

\usepackage{braket}
\usepackage{tikz}
\usepackage[normalem]{ulem}
\title{Quest for new physics using astrophysical neutrino flavor in IceCube}

\ShortTitle{Quest for new physics in IceCube}

\author{
The IceCube Collaboration\footnote{For collaboration list, see PoS(ICRC2019) 1177.}\\
{\itshape \href{http://icecube.wisc.edu/collaboration/authors/icrc19_icecube}{http://icecube.wisc.edu/collaboration/authors/icrc19\_icecube}}\\
E-mail: \email{k.r.h.farrag@qmul.ac.uk}
}

\abstract{
We have detected astrophysical neutrinos in IceCube that can be used to probe astrophysical sources at ultra high scales. Here we report a search for anomalous space time effects using astrophysical neutrino flavor data in IceCube. New effective operators are introduced to drive non-standard neutrino flavor mixing which modify the flavor ratios compared to standard cases. Using the High Energy Starting Events sample (HESE) 7.5-year data for this analysis, we found no evidence of such flavor anomalies. However, we are expecting to set limits from this new approach which goes far beyond any known techniques. Importantly, we achieve the necessary precision to probe new physics using neutrino flavor expected by Planck scale theories. \\

\vspace{4mm}
{\bfseries Corresponding authors:}
 Carlos A. Arg\"uelles$^{1}$, \speaker{Kareem Farrag}${}^{2,3}$,  Teppei Katori$^{2}$, Shivesh Mandalia$^{2}$\\
{$^{1}$ \itshape Department of Physics, Massachusetts Institute of Technology, Cambridge, MA 02139, U.S.A.}\\
{$^{2}$\itshape School of Physics and Astronomy, Queen Mary University of London, London, E1 4NS, U.K.}\\
{$^{3}$ \itshape Physics and Astronomy, University of Southampton, Southampton, SO17 1BJ, U.K.}
}

\FullConference{36th International Cosmic Ray Conference -ICRC2019-\\
		July 24th - August 1st, 2019\\
		Madison, WI, U.S.A.}

\definecolor{thegreen}{HTML}{00B110}
\definecolor{thepink}{HTML}{E71D73}
\definecolor{theblue}{HTML}{007DD2}
\definecolor{theorange}{HTML}{F39200}
\begin{document}
\section{Introduction}\label{sec:intro} 
Cosmic neutrinos offer us important insight in the search for extensions beyond the Standard Model with massive neutrinos, denoted here as the $\nu$SM. These neutrinos originate from extra-galactic astrophysical particle accelerators, carrying very high energies and traversing propagation distances which can exceed $O(100~\textup{Mpc})$.
Some of them arrive at the geographical South Pole of the Antarctic, where the IceCube Neutrino Observatory \cite{Aartsen:2016nxy} detects them in the High Energy Starting Event (HESE) sample \cite{Aartsen:2014gkd}. 
A span of 5,160 Digital Optical Modules (DOMs) lie beneath the ice capable of translating the Cherenkov light produced through charged particle radiation into signal. These charged particles are produced through the interaction of the incoming neutrino with the nucleons and electrons in the Antarctic ice sheet.

Efforts to extend the $\nu$SM are needed to accommodate experimentally observed phenomena, such as dark matter and gravity. Interestingly, quantum-gravity-motivated space-time effects (QG) are a frequent prediction near the Planckian regime $E_{pl} = \sqrt{\hbar c/G_{N}} \sim 10^{19}~ \textup{GeV}$. Both quantum field theory and general relativity work remarkably well at microscopic and macroscopic distance scales respectively, though a theory of quantum gravity is desirable since beyond $E_{pl}$ we expect quantum effects of gravity dominate. Furthermore, current unification theories such as string theory \cite{Kostelecky:1988zi,Mavromatos:2007xe} and modified gravity \cite{Kostelecky:2003fs,Myers:2003fd} come to the conclusion that QG is allowed within their frameworks as well as other anomalous space-time effects. Since neutrinos interact via the weak interaction and gravity due to their small masses, they can inform us about physics in regions of strong gravity throughout the cosmos, where gravitation might exhibit quantum behavior.

Several searches for new physics through atmospheric neutrino flavor composition have been conducted in the past, such as \cite{Abe:2014wla,Abbasi:2010kx,Aartsen:2017ibm}. Although, these are among the most sensitive spacetime tests on Earth, the sensitivity of this approach is limited by the Earth's size ($\sim$12,700~km) and the tail of the conventional atmospheric neutrino spectrum ($\sim$20~TeV). In the case of new physics searches via astrophysical neutrinos, their $O(100~\textup{Mpc})$ baselines and energies reaching up to $O(\textup{PeV})$ have already been used to test spacetime effects such as Lorentz symmetry violation (LV). One such terrestrial search of LV utilises the neutrino's time-of-flight (ToF). This test investigates whether light and neutrinos simultaneously emitted from a distant source arrive simultaneously on Earth. By comparing the difference in their arrival times, an incident neutrino's ToF is estimated and used to search for anomalous spacetime effects~\cite{Jacob:2006gn,Kostelecky:2011gq,Amelino-Camelia:2016ohi,Huang:2018ham,Ellis:2018ogq}. However, since we do not know precisely the emission time of neutrinos, nor the location of such sources, this test is experimentally challenging. On the other hand, it has been established that tests harnessing flavor information of the astrophysical neutrino flux can be used as a powerful probe for QG \cite{arguelles2015effect}. In this approach, we do not require emission time and source location information, but instead use flavor information from all astrophysical contributions in a simple effective operator framework.

\section{Analysis}\label{sec:analysis}

In this work, we perform the first QG test with astrophysical neutrino flavor data. We consider the flavor composition coming from diffuse astrophysical neutrinos via an effective operator framework as motivated by the Standard Model Extension (SME)~\cite{Kostelecky:2003fs}.
Given that neutrino oscillations occur through an effective Hamiltonian, we can encode the presence of new physics through a linear sum of effective operators with energy dimension $d$, which we can express as 

\begin{equation}
    H = \frac{1}{2E}UM^2 U^{\dagger} + \sum_{d>3}\frac{E^{d-3}}{\Lambda_d} \tilde{U}_d O_d \tilde{U}_d^{\dagger},
\end{equation}
where $U$ is the Pontecorvo-Maki-Nakagawa-Sakata matrix, $M^2$ denotes the neutrino squared mass matrix assuming neutrino Standard Model ($\nu$SM) paradigm \cite{Esteban:2018azc}, $\Lambda_d$ is the scale of new physics for a particular dimension $d$ operator, $O_d$ is a diagonal matrix and $\tilde{U}_d$ denotes the beyond the standard model (BSM) mixing matrix used to diagonalize each BSM operator respectively. Here, the first term is the vacuum Hamiltonian responsible for neutrino oscillations, written in the weak eigenstate basis. The other terms encode any new interactions, including neutrinos interacting with new space-time structure in the vacuum.  This effective Hamiltonian, regardless of including new interactions or not, causes neutrino mixings. Since neutrinos propagate cosmological distances, the phase information neutrinos carry is subsequently washed out. Hence, we can express neutrino mixing purely through an effective mixing matrix $V_d$ which diagonalizes the effective Hamiltonian.

In this analysis, we assume that one of the effective operators of a given dimension dominates the effective potential at a particular energy scale. This happens typically when new physics is at the order $\Lambda_d \sim E^{d-2}/M^2$ for some dimension $d$. By making this assumption, we can simplify the effective Hamiltonian we consider to have the following form, 

\begin{equation}
    H_d \sim \frac{1}{2E}UM^2 U^{\dagger} + \frac{E^{d-3}}{\Lambda_d} \tilde{U}_d O_d \tilde{U}_d^{\dagger}.
\end{equation}

It is reasonable to assume that each new physics operator will dominate at different energies because they correspond to different processes (e.g. scattering, decays etc.). These processes can be suppressed or enhanced by considering renormalization and thermal corrections at different scales. Hence, this motivates us to test each dimension operator one by one during the computation. 

Whilst the neutrino is produced and detected in its weak eigenstate, which we denote by Greek indices $\ket{\nu_\alpha}$, under propagation, the neutrino tracks eigenstates of the effective Hamiltonian, $H_d$, which we denote with Latin indices $\ket{\nu_i}$. We can, under a change of basis, diagonalize the effective Hamiltonian to construct the propagation eigenstate matrix $\Delta$ with respect to $H_d$ by the unitary transformation $V_d(E)$:

\begin{equation}\label{diagonalise}
    H_d \equiv V_d(E) \Delta(E) V_d(E)^{\dagger}.
\end{equation}

The matrices $V_d$ and $\Delta$ can then be computed using the Cardano equation~\cite{Abe:2014wla,Katori:2006mz}.
In neutrino oscillations with interactions, oscillation frequencies are defined up to the difference between the effective Hamiltonian eigenvalues, $\lambda_{i}(E)$, of the total Hamiltonian; namely $\Delta_{ij}(E) = \lambda_{i}(E) - \lambda_{j}(E)$. Similarly to the vacuum case, oscillation amplitudes are also defined by products of the effective mixing matrix $V_d$. We can therefore calculate the evolution a neutrino undergoes given this Hamiltonian with respect to its propagation. This is analogous to the case of standard neutrino oscillations and conversions through matter potentials within the $\nu$SM.

For all length scales and assuming the plane wave approximation for neutrinos, the oscillation probability due to the effective mixing matrix $V_d(E)$ as defined in Eq.~(\ref{diagonalise}) as
\begin{equation}
    P_{{\nu_\alpha}\rightarrow {\nu_\beta}}(L,E)= \delta_{\alpha\beta} - 4 \sum_{i>j}\textup{Re}\left(V^*_{\alpha i}V_{\beta i}V_{\alpha j}V_{\beta j}^{*}\right)\sin^2 \left(\frac{\Delta_{ij}L}{2}\right) + 2\sum_{i>j}\textup{Im}\left(V^*_{\alpha i}V_{\beta i}V_{\alpha j}V_{\beta j}^{*}\right)\sin \left(\Delta_{ij}L\right).
\end{equation}
In the limit that no BSM physics occurs, then this reduces to the standard oscillation formula with $\Delta_{ij}(E)\rightarrow \Delta m^2_{ij}/2E$ and $V_{ij}(E)\rightarrow U_{ij}$. Note, given the best fit for the atmospheric mass splitting $\Delta m^2_{\textup{atm}}\sim 2.3 \times 10^{-3}~\textup{eV}^2$, the oscillation length assuming a $\sim$PeV corresponds to $L_{\textup{osc}}\sim 0.0005$~pc. When neutrinos propagate longer than such scales ({\it e.g.}, the distance to TXS056+056 is $\sim$1.75 Gpc~\cite{IceCube:2018dnn}), the oscillatory features of their conversions are lost from averaging and decoherence. In such scenarios, the probability a neutrino converts according to effective mixing matrix $V$ can be written as

\begin{equation}
     \lim_{L\rightarrow \infty} P_{{\nu_\alpha}\rightarrow {\nu_\beta}}(L,E)=  \sum_{i} \left| V_{\alpha i}(E)\right|^2 \left|V_{\beta j}(E)\right|^2.
\end{equation}

The production of astrophysical neutrinos is consistent with the relevant UHECR production mechanisms and fluxes of high energy neutrino can be correlated with gamma ray data~\cite{ahlers2015high}. However, currently we cannot constrain the precise production mechanisms for high energy neutrinos.

\begin{figure}[!th]
    \centering
    \includegraphics[scale=0.3]{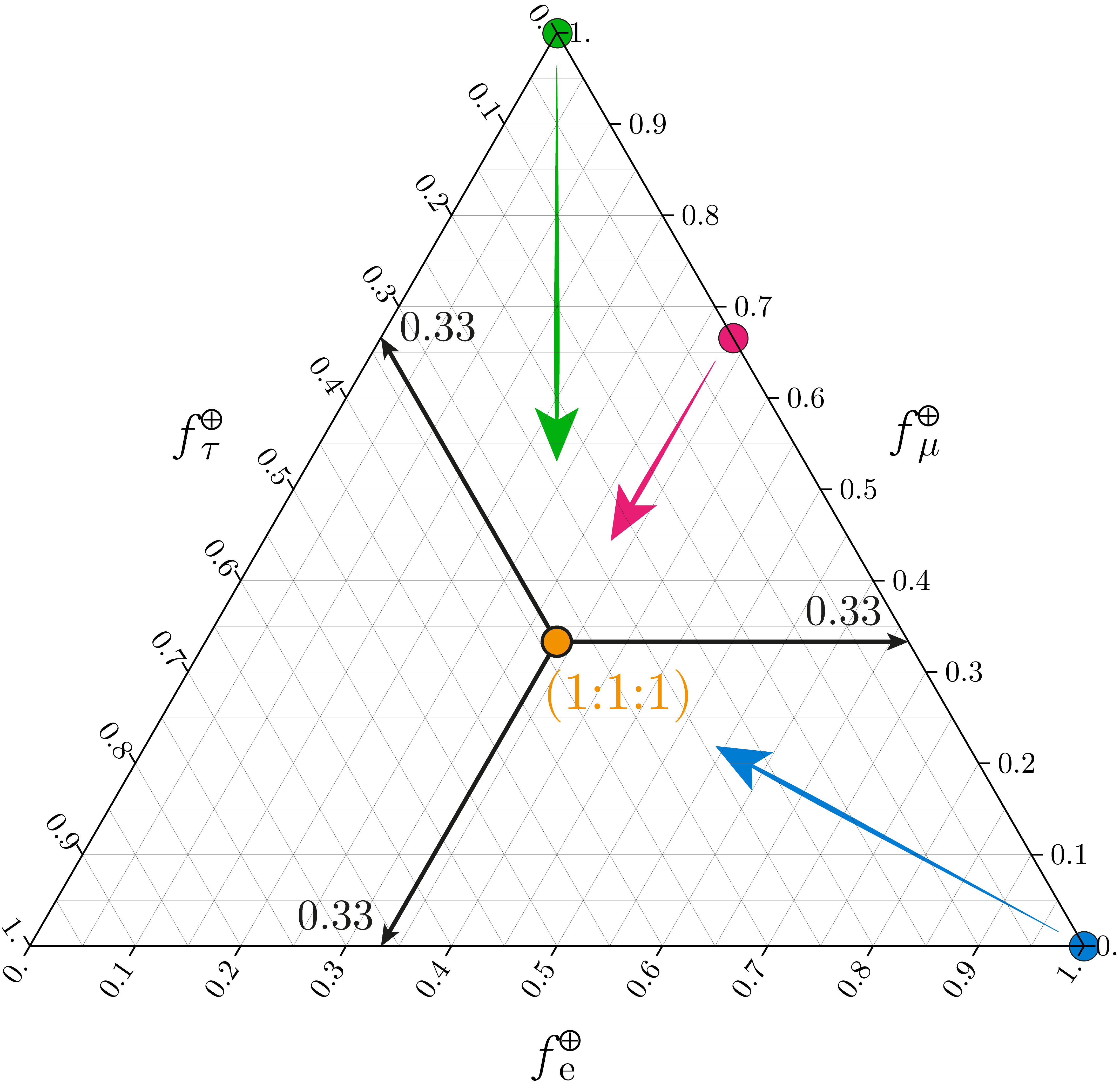}
    
    {\vspace{2pt}}
    \fbox{\begin{minipage}[t][0.75\height][c]{0.7\textwidth}
      \hfill
    \begin{tikzpicture}
    {\draw[black,thick,fill=thegreen] (0,0) circle (5pt) (0.2,0) node[anchor=west] {$(0:1:0)$};}
    \end{tikzpicture} \hfill
    \begin{tikzpicture}
    {\draw[black,thick,fill=thepink] (0,0) circle (5pt) (0.2,0) node[anchor=west] {$(1:2:0)$};}
    \end{tikzpicture}
    \hfill
    \begin{tikzpicture}
  {\draw[black,thick,fill=theblue] (0,0) circle (5pt) (0.2,0) node[anchor=west] {$(1:0:0)$};}
    \end{tikzpicture}
    \hfill
    \begin{tikzpicture}
    {\draw[black,thick,fill=theorange] (0,0) circle (5pt) (0.2,0) node[anchor=west] {$(1:1:1)$};}
    \end{tikzpicture}
    \hfill
    \end{minipage}
    }
    \caption{Figure shows an important representation of flavor ratio through the flavor triangle. 
    To read a point, one considers three lines subtended from the 
    point to the edges of the large triangle. These three lines are parallel to the triangle itself, and intersect each flavor axis at some given value. Note the line used to describe flavor $f_\alpha^\oplus$ always lies opposite the vertex $f_\alpha^\oplus=1$ and is parallel to the side $f_\alpha^\oplus=0$. This is illustrated with the orange point $(1:1:1)$. The green, magenta and blue points are the source ratios motivated by cosmic ray neutrino production. In the presence of neutrino oscillations or mixing effects, points lying towards the edge of the flavor triangle tend towards the central region, as indicated by the colored arrows.}
    \label{fig:triangle}
\end{figure}

We introduce $\phi_\alpha^S$ as the diffuse flux component for neutrinos of flavor $\alpha$ coming from such a source.
Three source composition models are predicted to dominate the flux: the production of neutrinos through pion decays which produces neutrinos in the ratio $(1:2:0)$; neutrino production via beta decay yielding a $(1:0:0)$ proportion; and through muon damping which produce a flavor composition of $(0:1:0)$. There is currently not a convincing mechanism able to produce a dominant astrophysical tau neutrino flux in $(0:0:1)$, though previous studies describe possible processes such as multiplication of $\nu_\tau$ yield through secondary decays \cite{Barish:2001iv,Weinstein:1993jp}.  Given some flavor ratio combination $(\frs)$ at the source, there are unitarity bounds on the available flavor ratios on the Earth~\cite{arguelles2015effect}. As a result, measuring the flavor ratio measurement in IceCube is a powerful mechanism to pin down possible production models. Assuming no new physics, all these models result in a terrestrial composition close to $\sim(1:1:1)$ on the Earth \cite{Aartsen:2014njl}. A well used tool in the literature for illustrating the allowed phase space of flavors at the Earth is the flavor triangle, as illustrated in Figure~\ref{fig:triangle}.


Given our flux compositions, we can compute the predicted flux ratio on Earth as
\begin{equation}
\phi_{\alpha, d}^{\oplus}=\frac{1}{\Delta E} \int_{\Delta E} \sum_{i, \beta}\left|V_{\alpha i, d}(E)\right|^{2}\left|V_{\beta i, d}(E)\right|^{2} \phi_{\beta}^S(E) d E.
\end{equation}

We can then normalise the ratio at the source by considering $f_{\alpha}^{S}= \phi_{\alpha}^{S}/(\sum_{\beta = e,\mu,\tau} \phi_{\beta}^{S})$ to represent our results via the flavor triangle. In this analysis, we wish to search for $\Lambda_d$ for any given dimension. Since we do not have precise forms of the BSM mixing matrix $\tilde{U}$, we must also sample these uniformly over $SU(3)$ to prevent bias in the choice of $U$. 

This is done using the Haar measure, which for 3x3 matrices parametrized in terms of BSM mixing angles $\tilde{\theta}_{12}, \tilde{\theta}_{13}, \tilde{\theta}_{23}$ and the BSM CP phase $\tilde{\delta}$ is given by
\begin{equation}
d\tilde{U} = d(\sin^2\tilde{\theta}_{12})\wedge d(\cos^4\tilde{\theta}_{13}) \wedge d(\sin^2\tilde{\theta}_{23}) \wedge d(\tilde{\delta}).
\end{equation}

In order to incorporate systematic errors effectively, we use Bayesian analysis utilizing Markov Chain Monte Carlo (MCMC).~\cite{Feroz:2008xx}. Sampling over the joint posterior distribution, MCMC allows us to explore parameter spaces in moderately high dimensions. In terms of the likelihoods, given a null hypothesis $\Theta_0$ and alternative hypothesis $\Theta_1$ and a data sample $D$, the Bayes factor $B$ can be expressed as
\begin{equation}
B=\frac{P(D|\Theta_{1})}{P(D|\Theta_{0})}, 
\end{equation}
where the numerator and denominator are the Bayesian likelihoods for given hypothesis $P(D|\Theta)$, respectively. $B>1$ implies a preference to the alternative hypothesis $\Theta_1$, whereas $B<1$ is favoured by the null $\Theta_0$. A convention is chosen to describe the strength of favorability between the two hypotheses. We choose to use the Jeffreys' scale as defined in Table~\ref{tab:jeffreys}~\cite{jeffreys1998theory}
\begin{figure}[!h]
\begin{center}
\begin{tabular}{c|c}
\hline
Bayes factor, $B$ & Inference Convention \\
\hline
    $B <1$ & $\Theta_0$ favoured \\
    $1<B<10^{1/2}$ & $\Theta_1$ barely favoured\\
    $10^{1/2}<B<10^{1}$ & $\Theta_1$ substantially favoured\\
    $10^{1}<B<10^{3/2}$ & $\Theta_1$ strongly favoured\\
    $10^{3/2}<B<10^{2}$ & $\Theta_1$ very strongly favoured\\
    $10^{2}<B$ & $\Theta_1$ decisively favoured\\
    \hline
\end{tabular}
\end{center}
\caption{List of Bayes factors and their convention according to Jeffreys' scale}
\label{tab:jeffreys}
\end{figure}

\subsection{The High Energy Starting Event (HESE) 7-year Data Sample}\label{sec:hese}

Our analysis makes use of the 7 year High Energy Starting Events (HESE) in IceCube \cite{Aartsen:2014gkd}. We define the veto region by the outer layer of DOMs in the detector. Inside of this region is the fiducial volume of interest, where we select events if they start inside and produce more than 6000 photoelectrons. We further select 60~TeV as the minimum reconstructed energy to ensure a low contamination in the sample from atmospheric events. In the event of charged-current interactions, a neutrino of a given flavor produces charged leptons of the same flavor. The interaction of each flavor can generally be classified into signature morphologies in IceCube. For instance, $e^{-}$ ($e^{+}$) produced through $\nu_e$ ($\bar{\nu}_e$) charged-current interactions can produce a isotropic emission of photons. Similarly $\nu_\mu$ (muon $\bar{\nu}_\mu$) can produce $\mu^{-}$ ($\mu^{+}$) that develop track-like signatures in the ice. Finally, $\nu_\tau$ ($\bar{\nu}_{\tau}$) can at very high energies produce a boosted $\tau^{-}$ ($\tau^{+}$) that deposits electromagnetic and hadronic showers separated by the distance the $\tau$ decays. Note, IceCube cannot distinguish charged leptons from antileptons, and in the following sections we refer neutrinos to include both neutrinos and antineutrinos.

The 7-year HESE sample include 102 events in total, with both cascade and track morphologies; 60 of these events occur above 60 TeV where we are most sensitive to astrophysical neutrinos.  

\section{Search of new physics through astrophysical neutrino flavor}\label{sec:result}

As part of our investigation, we follow the previous HESE flavor analysis \cite{Aartsen:2014gkd} by binning logarithmically over 20 energy bins from 60 TeV to 10 PeV. We also split zenith angle over 10 bins for $-1<\cos\theta<1$ assuming the flux is isotropic along azimuth angle. We incorporate simulations of the fluxes and cross sections from \cite{Honda:2006qj,Bhattacharya:2015jpa,CooperSarkar:2011pa}.  

We introduce the following systematic uncertainties in this analysis. These include flux normalizations for the atmospheric muon, and conventional, prompt, and astrophysical neutrinos. We also assume a single power law for the astrophysical neutrino spectrum. We then include the 6 oscillation parameters with uncertainties i.e., the SM mixing angles, phase and square masses. \cite{Esteban:2018azc}. This leads to a total of 11 nuisance parameters. 

Here we present some examples from this analysis~\cite{Katori:2019xpc}. The following plots show how the Bayes factor changes with respect to the scale of new physics. We assume the $d=6$ dominant case with source flavor ratio of $(1:0:0)$ and $(0:1:0)$, respectively. The horizontal axis refers to the new physics scale and the vertical axis is the value of the Bayes factor. The red horizontal line is plotted at the threshold value of $B=10^{3/2}$ where we define the strong favorability for new physics convention. We note that the Bayes factor in both scenarios pass the threshold value below the range $ \Lambda_{6}^{-1}\lesssim 10^{-45}~\textup{GeV}^{-2}$. This corresponds to an energy scale of new physics around $10^{22}~\textup{GeV}\lesssim \sqrt{\Lambda_{6}}$. Several assumptions are required, but using this technique we are able to constrain new physics expected around the Planckian regime. Since we use model independent effective operators in our analysis, our constraints can further be recast in terms
of other BSM physics scenarios including LV~\cite{Kostelecky:2011gq}, long range forces~\cite{Bustamante:2018mzu}, neutrino-dark energy coupling~\cite{Klop:2017dim}, neutrino-dark matter coupling\cite{Farzan:2018pnk}, and many others~\cite{Rasmussen:2017ert}.

\begin{figure}[!h]\label{fig:bayes}
    \centering
    \includegraphics[scale=0.4]{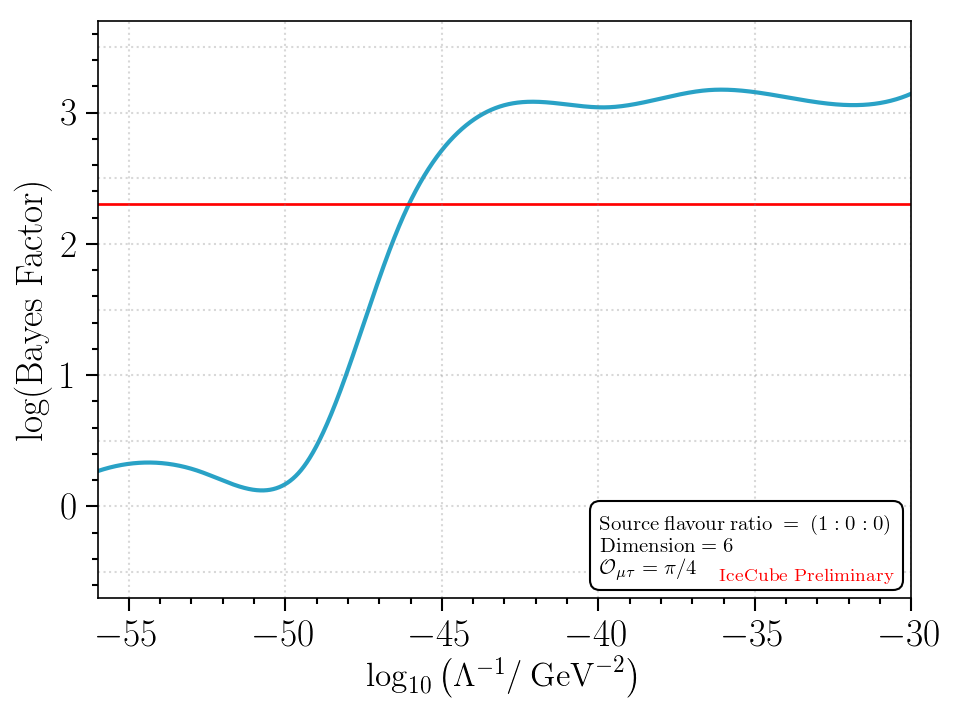}
    \includegraphics[scale=0.4]{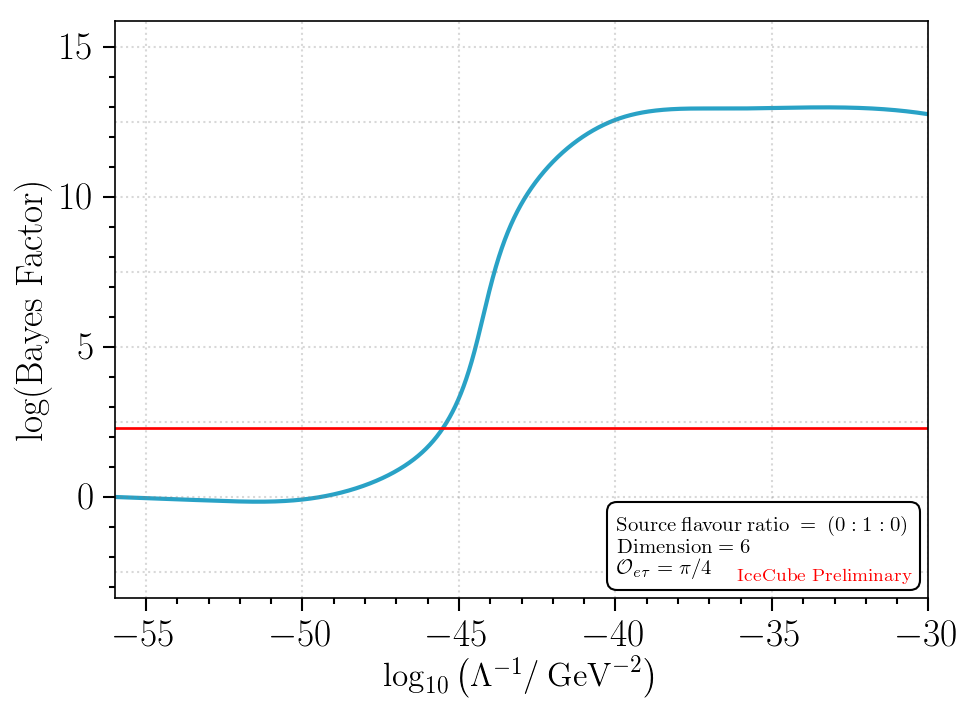}
    \caption{Plot of the Bayes factor as a function of the scale of new physics. Left shows the case assuming a source flavor ratio of $(1:0:0)$ and right $(0:1:0)$ respectively. }
    \label{fig:my_label}
\end{figure}
\section{Conclusion}\label{sec:conclusion}
We have performed the search for new physics using the HESE 7-year data flavor information with MCMC sampling. We have shown how neutrinos can be a powerful tool in constraining new physics. Given source flavor ratios of $(1:0:0)$ and $(0:1:0)$, we achieved $\sqrt{\Lambda_6}\gtrsim 10^{22}$~GeV. We start to explore the phase space where we anticipate quantum gravity-motivated new physics. In the future, higher statistics with regards to astrophysical neutrino flavor data will improve our ability to probe new physics. The next generation experiments, such as IceCube-Gen2 \cite{Aartsen:2014njl}, will have an order of magnitude more effective area with improved optical sensors. Such detectors will begin to investigate Planck scale physics at an unprecedented level, as well as bolster new physics searches in the landscape of multimessenger astronomy.
 
{\color{red}}
\bibliographystyle{ICRC}
\bibliography{references}

%

\end{document}